\documentclass[preprint,review,final]{elsarticle}
\usepackage{amsmath}
\usepackage{amsfonts}
\usepackage{gensymb}
\usepackage{lineno,hyperref}
\usepackage[dvipsnames]{xcolor}
\usepackage{subcaption}
\usepackage{xcolor}
\usepackage[utf8x]{inputenc}
\usepackage{chngcntr}
\counterwithout{figure}{subsection}

\usepackage{changes}
\usepackage{float}
\usepackage{mathtools}
\usepackage[linesnumbered, ruled, vlined]{algorithm2e}
\usepackage[shortcuts]{extdash}
\usepackage{soul}
\journal{Nuclear Inst. and Methods in Physics Research, A}

\def\bfR{{\mathbf{R}}}
\journal{Nuclear Inst. and Methods in Physics Research, A}

\newcommand{\bfy}{\textbf{y}}

\newcommand{\bphi}{{\boldsymbol \phi}}

\newcommand{\transp}{^T}

\makeatletter
\def\ps@pprintTitle{
 \def\@oddfoot{}%
 \let\@evenfoot\@oddfoot}
\makeatother

\begin{document}


\begin{frontmatter}

\title{Characterization of stilbene\-/d$_{12}$ for neutron spectroscopy without time of flight}


\author[mymainaddress]{N. Gaughan}

\author[mymainaddress]{J. Zhou}
\author[mysecondaryaddress]{F. D. Becchetti}
\author[mysecondaryaddress]{R. O. Torres-Isea}
\author[mythirdaddress]{M. Febbraro}
\author[myfourthaddress]{N. Zaitseva}
\author[mymainaddress]{A. Di Fulvio\corref{mycorrespondingauthor}}
\cortext[mycorrespondingauthor]{Corresponding author.}
\ead{difulvio@illinois.edu}

\address[mymainaddress]{Department of Nuclear, Plasma, and Radiological Engineering, \\University of Illinois, Urbana-Champaign,
                        \\ 104 South Wright Street, Urbana, IL 61801, United
                        States}
\address[mysecondaryaddress]{Department of Physics, University of Michigan, Ann Arbor, MI, US}
\address[mythirdaddress]{Oak Ridge National Laboratory, Oak Ridge, TN, US}
\address[myfourthaddress]{Lawrence Livermore National Laboratory, Livermore, CA, US}

\begin{abstract}
We have experimentally characterized the light\-/output response of a deuterated trans-stilbene (stilbene\-/d$_{12}$) crystal to quasi-monoenergetic neutrons in the 0.8 to 4.4 MeV energy range. These data allowed us to perform neutron spectroscopy measurements of a DT 14.1 MeV source and a $^{239}$PuBe source by unfolding the impinging neutron spectrum from the measured light\-/output response. The stilbene\-/d$_{12}$ outperforms a $^1$H$\-/$stilbene of similar size when comparing the shape of the unfolded spectra and the reference ones. These results confirm the viability of non-hygroscopic stilbene\-/d$_{12}$ crystal for direct neutron spectroscopy without need for time-of-flight measurements. This capability makes stilbene\-/d$_{12}$ a well suited detector for fast\-/neutron spectroscopy in many applications including nuclear reaction studies, radiation protection, nuclear non-proliferation, and space travel.
\end{abstract}


\end{frontmatter}


\section{Introduction}

Elastic neutron scattering on light nuclei is the main interaction that enables neutron detection in organic scintillators in the fast neutron energy range (100~keV - 20~MeV). The energy deposited within the detector volume is absorbed and subsequently emitted in the form of photons in the blue-UV range and converted into current pulses by light-readout devices, such as photo-multiplier vacuum tubes (PMTs). The energy deposited, and hence the light output readout signal, depends on the impinging neutron energy and the neutron scattering angle in the scintillator. Therefore, organic scintillators' light output response provides a measurement of the energy of the impinging neutrons. However, performing direct neutron spectroscopy with hydrogen-based scintillators is a challenging task because the neutron scattering reaction on protons is primarily isotropic (in the 0$^o$-90$^o$ neutron scattering angle $\theta_n$ range). Therefore, at a given impinging neutron energy $E_0$, the energy deposited in the scintillator by proton recoils is uniformly distributed from zero ($\theta_n = 0\degree$, proton recoil scattering angle $\theta_p = 90\degree$) to $E_0$ ($\theta_n = 90\degree$, $\theta_p = 0\degree$). Hydrogen-based scintillators hence yield smooth light output spectra from which it is challenging to deconvolve the energy of the interacting neutrons, especially in the case of poly-energetic fields. Conversely, neutron elastic scattering on deuterium is non-isotropic, with a higher cross section for back-scattered neutrons than forward-scattered ones. As a result, the light output response of deuterated scintillators to monoenergetic neutrons ($E_0$) shows a peak-like structure corresponding to the maximum energy deposited by deuteron recoils, i.e., $8/9\times E_0$ \cite{Becchetti_deutscint}. 
This property results in the favorable spectroscopy capabilities of deuterated scintillators. Spectroscopy applications have already been demonstrated for liquid deuterated benzene \cite{FEBBRARO2015184} and xylene \cite{DiFulvio_dXyl},\cite{Becchetti_dXyl}. In this work, we characterized the neutron light-output response, pulse shape discrimination (PSD), and spectroscopy capabilities of deuterated crystalline trans-stilbene (stilbene-d$_{12}$), a recently developed non-hygroscopic, solid-state, scintillating crystal \cite{Becchetti_dsb}. 
The 32~cm$^3$ volume stilbene-d$_{12}$ characterized in this work is a 3.55~cm thick hexagonal prism, which was grown at Lawrence Livermore National Laboratory through a solution growth method \cite{CARMAN201851}. 

Neutron spectroscopy is needed in several applications. Nuclear power plants and other nuclear facilities must monitor neutron fields for radiation protection to validate neutron transport models \cite{linacs}, and monitor plutonium production. In space flight, cosmic rays interact with spacecraft materials to produce secondary radiation, including neutrons. An accurate characterization of the neutron spectrum is necessary for the radiation protection of the astronauts \cite{Kinnison}. In nuclear physics, scintillation detectors are used to study nuclear reactions involving neutrons \cite{Bogart}, and deuterated scintillators can perform this task without the need for time-of-flight (TOF) measurements, allowing for use of high-intensity DC accelerators \cite{Febbraro2013}.

\section{Methods}
This section introduces the methods used to exploit the stilbene-d$_{12}$ PSD capability and to characterize stilbene-d$_{12}$'s response to quasi-monoenergetic neutrons. We also briefly introduce the spectrum unfolding algorithms and the metrics used to compare the spectroscopy capability of stilbene-d$_{12}$ to $^1$H$\-/$stilbene.

\subsection{Pulse Shape Discrimination} \label{PSD}
\begin{figure}[h!]
    \centering
    \captionsetup{justification=centering}
    \includegraphics[width=100mm]{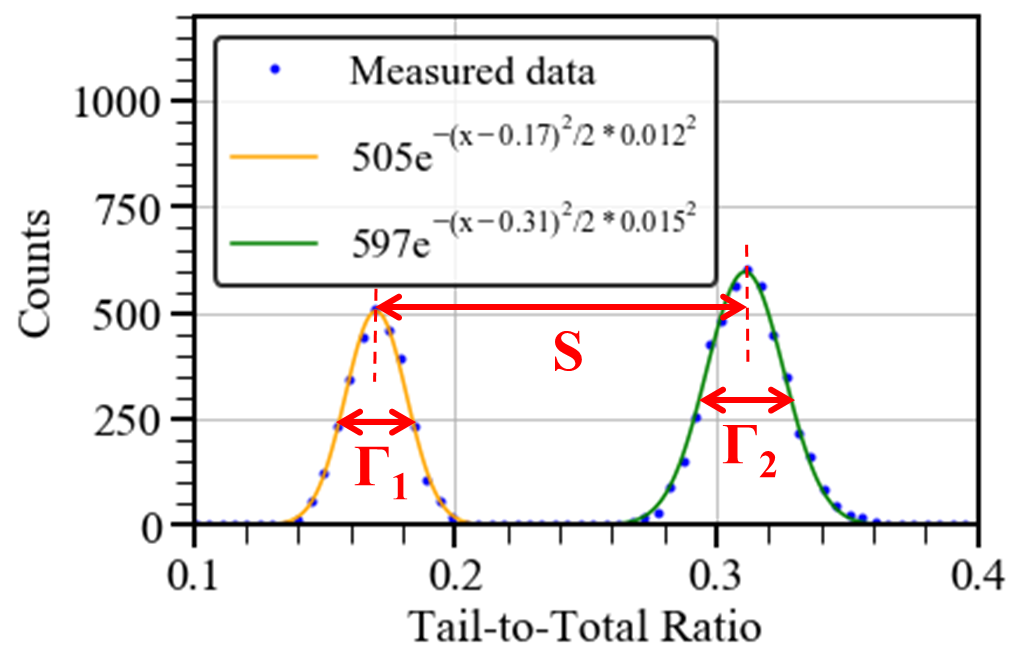}
    \caption{Distribution of $^{239}$PuBe counts in terms of tail-to-total ratio, in the 710-822 keVee light output interval. The gamma-ray peak is centered at 0.17 and the neutron peak is centered at 0.31.}
    \label{slice}
\end{figure}

Stilbene-d$_{12}$ is a PSD-capable detector. PSD enables the discrimination between pulses produced by different types of radiation based on their shape. PSD needs to be performed to select neutron pulses prior to neutron spectroscopy. 
In low-Z organic scintillators, gamma rays are detected through the recoil electrons produced by Compton scattering interactions. 
Ionizing radiation interactions are followed by molecular vibrations to the singlet or triplet states, whose decay can emit scintillating photons. Recoil electrons produce excited singlet states in the surrounding molecules, which emit prompt fluorescent light by decaying to the singlet ground states ($S_1 \rightarrow S_0$). Recoil protons and deuterons, produced by neutron scattering, exhibit a higher energy loss, dE/dx (linear energy transfer, LET), than electrons of the same energy, producing a higher ionization density along their path. The higher the LET, the higher the rate of formation of long-lived triplet states ($T_1$), which can undergo triplet-triplet annihilation (TTA) and yield a ground $S_0$ and an excited $S_1$ singlet state, which in turn decays with the emission of a delayed photon \cite{Horrocks}. This process is referred to as pyrene-type (p-type) delayed fluorescence and has longer characteristic times of the order of tens of ns, compared to the prompt fluorescence (1–2~ns) while maintaining the same spectral response. The overall effect of these processes is that a neutron pulse exhibits a more prominent delayed fluorescence component than a gamma-ray pulse, for a given energy deposited in the crystal. 
We used the charge-integration method \cite{POLACK2015253} with a fast pulse digitizer to discriminate neutron pulses from gamma-ray pulses. We calculated the integral of each pulse, $I_{total}$, and the integral of the tail portion of the pulse, $I_{tail}$, to capture the delayed fluorescence component and then calculated the tail-to-total ratio as $I_{tail}/I_{total}$. Neutron pulses are expected to exhibit a higher tail-to-total ratio than gamma-ray pulses.
The pulse tail is defined as starting $t_{start}$~ns after the pulse maximum. The total integral of the pulse is calculated over 374~ns, starting 4 ns before the time stamp of the pulse maximum.
Fig. \ref{slice} shows the distribution in tail-to-total ratio for pulses from a $^{239}$PuBe source (strength 1.3$\times$10$^6$ n/s, June 1958) with $t_{start}$ = 14ns, where the neutron and gamma-ray distributions in stilbene-d$_{12}$ can be clearly distinguished.

\begin{figure}[h!]
    \centering
    \captionsetup{justification=centering}
    \includegraphics[width=100mm]{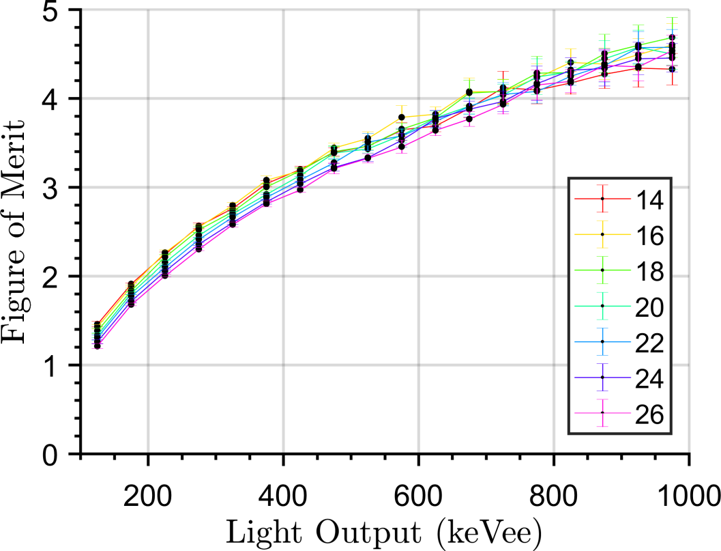}
    \caption{Stilbene-d$_{12}$ figures of merit as a function of light output for different values of $t_{start}$ in ns.}
    \label{FOM_fig}
\end{figure}

A figure of merit (FOM) is often used as a metric of the PSD capability of a scintillator and was used to optimize $t_{start}$. The FOM increases with greater spacing between the neutron and gamma distributions and with decreased full-width at half maximum (FWHM) of the distributions. The FOM is calculated using Equation \ref{FOM_eq} \cite{Taggart}.
\begin{equation}\label{FOM_eq}
    FOM = \frac{S}{\Gamma_1 + \Gamma_2}
\end{equation}
As shown in Fig.\ref{slice}, $S$ is the distance between the maximum values of the neutron and the gamma-ray distributions. $\Gamma_1$ and $\Gamma_2$ are the FWHMs of the gamma-ray and neutron distribution, respectively. We calculated the FOM for different $t_{start}$ values and selected $t_{start}$ = 14 ns because it yields the highest FOM at low light output values (Fig. \ref{FOM_fig}).
A $t_{start}$ and total integration gate of 22~ns and 374~ns, respectively, were selected to perform PSD of $^1$H$\-/$stilbene pulses after following a similar optimization procedure as the one described for stilbene-d$_{12}$.

\subsection{Time-of-Flight Technique}
\begin{figure}[h!]
    \centering
    \captionsetup{justification=centering}
    \includegraphics[width=125mm]{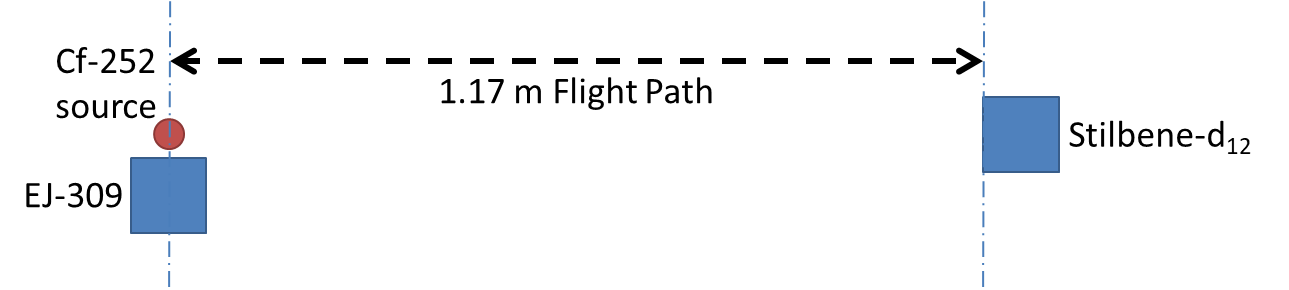}
    \caption{Schematic of the experimental setup used to determine the neutron light output response of the stilbene\-/d$_{12}$ detector through TOF.}
    \label{setup}
\end{figure}

The TOF technique was used to determine the stilbene\-/d$_{12}$ response to quasi-monoenergetic neutrons. We calculated the energy of the neutrons impinging on the detector using Equation \ref{neut_E}, where $E$ is the neutron energy, $m$ is the neutron mass, $c$ is the speed of light, and $v$ is the neutron speed. The latter is given by the ratio between the known travel distance and the measured travel time. 
\begin{equation}\label{neut_E}
    E = mc\textsuperscript{2}\left[\frac{1}{\sqrt{1-\frac{v^2}{c^2}}}-1\right]\approx\frac{1}{2}mv^2 \qquad v<<c
\end{equation}
The TOF measurement encompasses two detectors, a fission gamma ray \textit{start} detector placed next to the source, and the stilbene\-/d$_{12}$, \textit{stop} detector to be characterized, 1.17~m away from the source. This distance was chosen to obtain at least 10,000 counts in each energy bin while keeping the measurement time below 72 hours. In our case, the \textit{ start} detector was a 5.08 cm diam. by 5.08 cm length EJ-309 liquid scintillator. Fig. \ref{setup} shows a schematic of the experimental setup. We used a $^{252}$Cf spontaneous fission source with a 1.16$\times 10^7 \pm 5.8\times10^5$  neutrons/s source strength. Gamma rays emitted by fission simultaneously with neutrons were detected and used as the trigger signal to start the travel time measurement, stopped by the corresponding fission neutron detection event in the stilbene\-/d$_{12}$. This procedure allowed us to measure the TOF corresponding to each neutron detection event in the stilbene\-/d$_{12}$ and hence to calculate the energy of each impinging neutron that resulted in a detected pulse in the stilbene\-/d$_{12}$. 

We selected gamma-ray pulses in the start detector and neutron pulses in the stop detector through PSD.   
The timing of each detected event was determined through a digital constant fraction discrimination (CFD) algorithm, with an attenuation factor of 50\% fraction and a delay of 6 ns \cite{Ming}. CFD yields a bipolar pulse, whose time stamp is determined through linear interpolation of the zero-crossing region, between the sample before and after the zero crossing. The time distribution of the gamma-ray coincidence events was also measured and used to correct for the timing offset due to the front-end electronics and different cable lengths. 
The pulse integral distribution (PID) detected from the stilbene\-/d$_{12}$ detector was calibrated by measuring the Compton edge of a 37000-Bq (1 $\mu$Ci) $^{137}$Cs source, and the linear calibration coefficient was determined to be 5.53~MeVee/V. Detector pulses were acquired using a CAEN DT5730 digitizer, which features a sampling frequency of 500 MS/s, a 14-bit amplitude resolution, and a 2-Volt input range. 
The neutron pulses were sorted and binned based on the time of flight. 
The lowest neutron energy that can be accurately detected with this setup is 800 keV, with $\pm$ 22 keV uncertainty. This energy corresponds to a flight time of 94.6 ns. The highest energy bin is centered at 4400 keV, with $\pm285.3$ keV uncertainty, corresponding to a flight time of 40 ns. 
An analogous data collection procedure was repeated with a shadow bar in front of the stop detector to measure the contribution from neutrons that scattered off the floor and walls before reaching the stop detector. A 60.96 cm long and 4.7 cm diameter polyethylene cylinder was used as shadow bar. The shadow-bar data were then subtracted from the bare detector measurement to obtain the TOF distribution due only to the neutrons with a direct source-to-detector path. 

The resulting net detector light-output response to the quasi-monoenergetic neutron groups is a response matrix that in principle extends from zero to 8/9 of the incident neutron energy. 
The corresponding light output was then used to determine the stilbene\-/d$_{12}$ response to 
quasi-monenergetic neutrons in the 0.8 - 4.4 MeV energy range. The specific purpose of this work was to measure the stilbene\-/d$_{12}$ light output response to quasi-monoenergetic neutrons and use it to simulate the stilbene\-/d$_{12}$ response matrix in a wider energy range (0.1 - 20 MeV) in order to perform neutron spectrum unfolding, as explained in Section \ref{unfolding}.

\subsection{Unfolding Algorithm and Spectroscopy Metrics}\label{unfolding}
The neutron spectrum of a neutron source \textbf{$\phi(E)$} measured by an organic scintillator can be derived knowing the light-output response of the detector \textbf{$y(E')$} and its response matrix [$R(E',E)$)] to monoenergetic neutrons. $R(E',E_0)$ is the light output spectrum (with $E'$ in eVee) in response to a monoenergetic neutron of energy $E_0$. The neutron spectrum is related to the response matrix and the detector light-output readout by the Fredholm integral Equation (\ref{general}):
\begin{equation} \label{general}
y(E')=\int_0^{\infty}R(E',E)\phi(E)dE.
\end{equation}

Equation \eqref{general} can be approximated by the following linear equation:
\begin{equation} \label{discr}
\bfy  \approx \bfR \bphi,
\end{equation}
where $\bphi = [\phi_1,\ldots,\phi_N]\transp \in \mathbb{R}_{+}^N$ denotes the neutron spectrum discretized over $N$ energy bins, $\bfy=[y_1,\ldots,y_M]\transp \in \mathbb{R}_{+}^M$ is light output spectrum discretized over $M$ bins and $\bfR$ is the $M \times N$ response matrix of the detector. 
We used an iterative Bayesian unfolding method to derive the neutron spectrum from the measured light\-/output spectrum. In each iteration, the spectrum is found by combining the guess (obtained in the previous iteration) with the smoothing parameter, $\delta$. This process is described mathematically in Equation \ref{unfold_eqn} and further detailed in our previous work \cite{unfolding}. 

\begin{equation}\label{unfold_eqn}
f( \boldsymbol {{\phi }},\delta | {\mathbf {y}})=\frac {f( {\mathbf {y}}| \boldsymbol {{\phi }})f( \boldsymbol {{\phi }}|\delta )f(\delta )}{f( {\mathbf {y}})}\propto f( {\mathbf {y}}| \boldsymbol {{\phi }})f( \boldsymbol {{\phi }}|\delta )f(\delta )
\end{equation}

In Equation \ref{unfold_eqn}, $f(\mathbf{y}|\phi)$ is the probability of observing the light output $\mathbf{y}$ from the incident spectrum $\phi$, $f(\phi|\delta)$ is the prior distribution, and $f(\phi, \delta|\mathbf{y})$ is the posterior distribution. For the first iteration of the algorithm, we set $\phi^{(0)}=1$ (the initial guess). A Markov-Chain Monte Carlo (MCMC) technique was used to sample the neutron spectrum, approximated as the posterior mean of $\phi$ (Equation \ref{conf_int}). 
\begin{equation}\label{conf_int}
  \hat{\bphi}=1/(N_{iter}−N_{bi}) \Sigma^{N_{iter}}_{k=N_{bi}+1}\phi^{(k)}  
\end{equation}

In Equation \ref{conf_int}, $N_{iter}$ is the number of iterations, $N_{bi}$ is the number of initial burn-in iterations, and \textit{k} is the iteration number. The marginal posterior mean $\hat\phi$ is approximated by averaging the generated spectra after having removed the first $N_{bi}$ iterations, which correspond to the burn-in period of the sampler, set to 20\% of $N_{iter}$. The marginal 95\% credible interval (CI) for each element of $\hat\phi$ is
also calculated as the quantile of the elements in the sampler for the 95\% cumulative probability. 

The iterations ended when the absolute value of the relative difference between the measured light output and the convolution between the estimated neutron spectrum and the detector's response matrix was lower than 2\%, which corresponded to approximately 7000 iterations in the cases analyzed.

We unfolded the spectra of a $^{239}$PuBe source ($1.3\times10^6$ n/s, June 1958) and a D-T source (14.1 MeV neutrons, $\approx10^8$ n/s) from the light output measured by the stilbene\-/d$_{12}$ and a $^1$H$\-/$stilbene of similar size, shape, and growth method as the stilbene\-/d$_{12}$ \cite{Becchetti_dsb}.
The response matrices of the two detectors were simulated using MCNPX-PoliMi \cite{PoliMi} and MPPost using the light output coefficients detailed in Section \ref{respMatr}.

We calculated the Spectral Angle Mapper (SAM) (Equation \ref{eq:SAM} \cite{Keshava2002}) between the unfolded spectrum ($\bf\hat{\bphi}$) and a reference spectrum ($\bphi$) to compare the spectroscopy capability of the stilbene\-/d$_{12}$ with $^1$H$\-/$stilbene. The ground truth neutron spectra and response matrix may have different neutron energy resolutions. Therefore, we adopted the SAM as opposed to standard Mean Square Error (MSE) because the SAM is scale-invariant.
The SAM measures the spectral angle between $\bphi$ and $\hat{\bphi}$, which is small when $\bphi$ and $\hat{\bphi}$ have similar shapes.  
\begin{eqnarray}
\label{eq:SAM}
SAM(\bphi,\hat{\bphi})=\arccos \left(\frac{\bphi\transp \bf\hat{\bphi}}{||\bphi||_2||\hat{\bphi}||_2}\right). 
\end{eqnarray}

\section{Experimental Results}
\subsection{PSD Capability}

\begin{figure}[h!]
    \centering
    \captionsetup{justification=centering}
    \includegraphics[width=120mm]{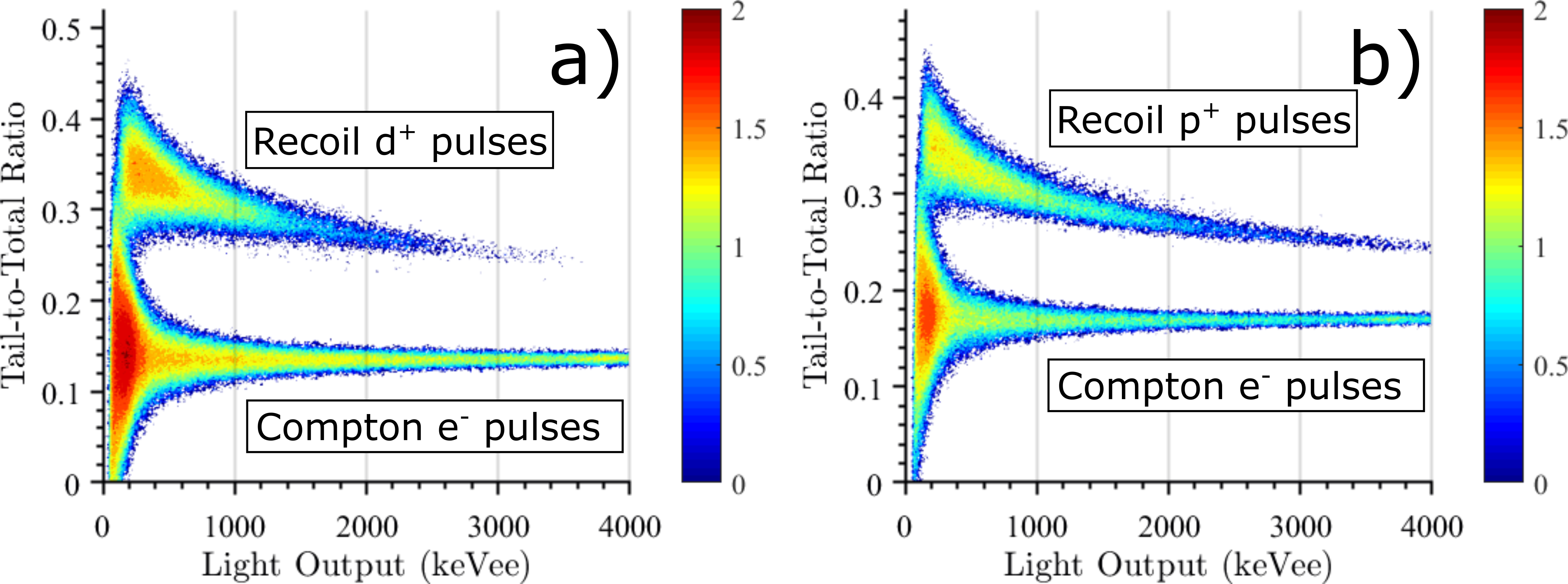}
    \caption{PSD of (a) stilbene\-/d$_{12}$ and (b) $^1$H$\-/$stilbene for a $^{239}$PuBe source. The colorbar indicates the number of pulses in a logarithmic scale.}
    \label{PSD_2plots}
\end{figure}

\begin{figure}[h!]
    \centering
    \captionsetup{justification=centering}
    \includegraphics[width=100mm]{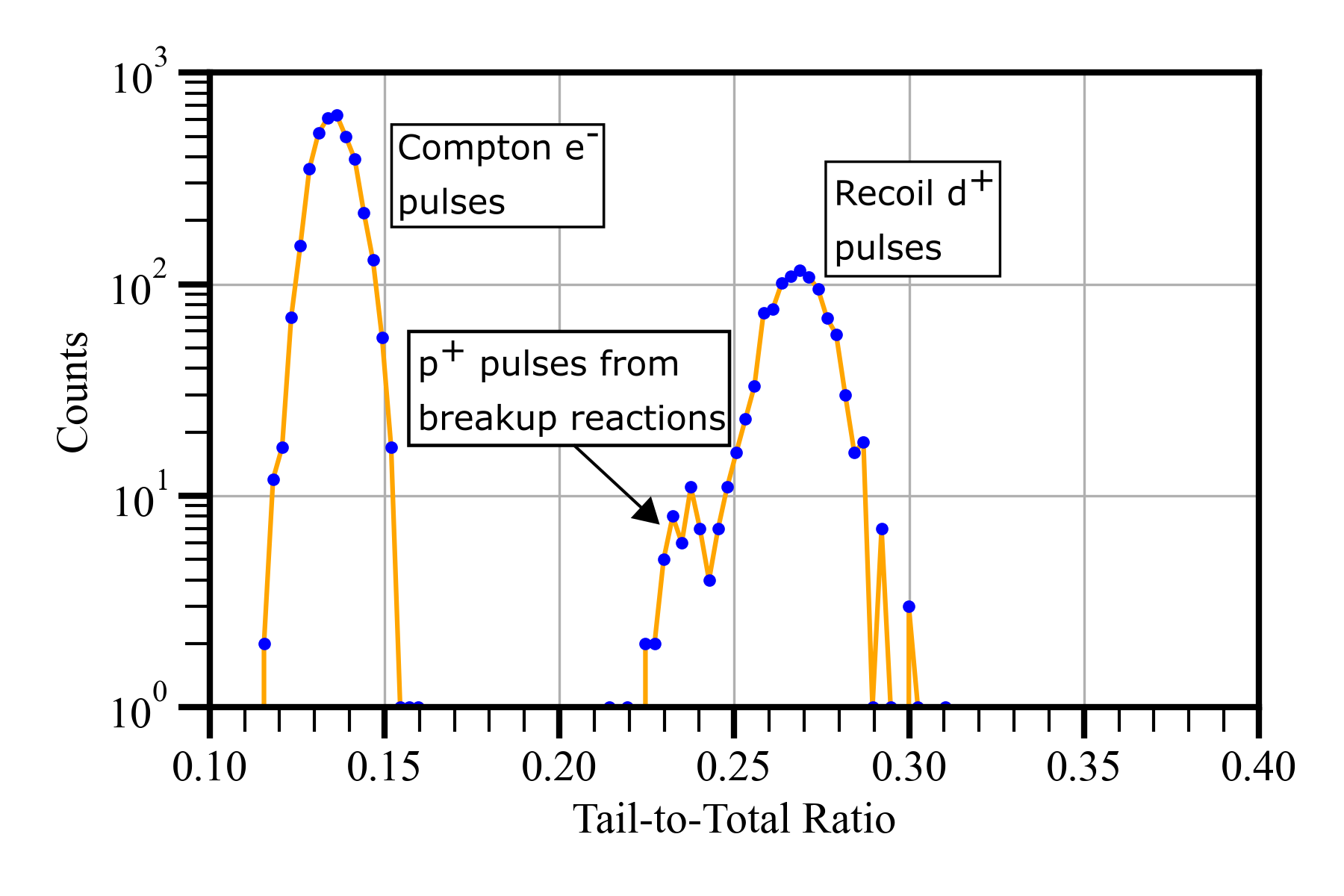}
    \caption{Slice of the PSD plot from the $^{239}$PuBe source shown in Fig. \ref{PSD_2plots}, from 2025 keVee to 2292 keVee. A small peak from protons is visible next to the deuteron recoil peak.}
    \label{PuBe_slice}
\end{figure}

\begin{figure}[h!]
    \centering
    \captionsetup{justification=centering}
    \includegraphics[width=80mm]{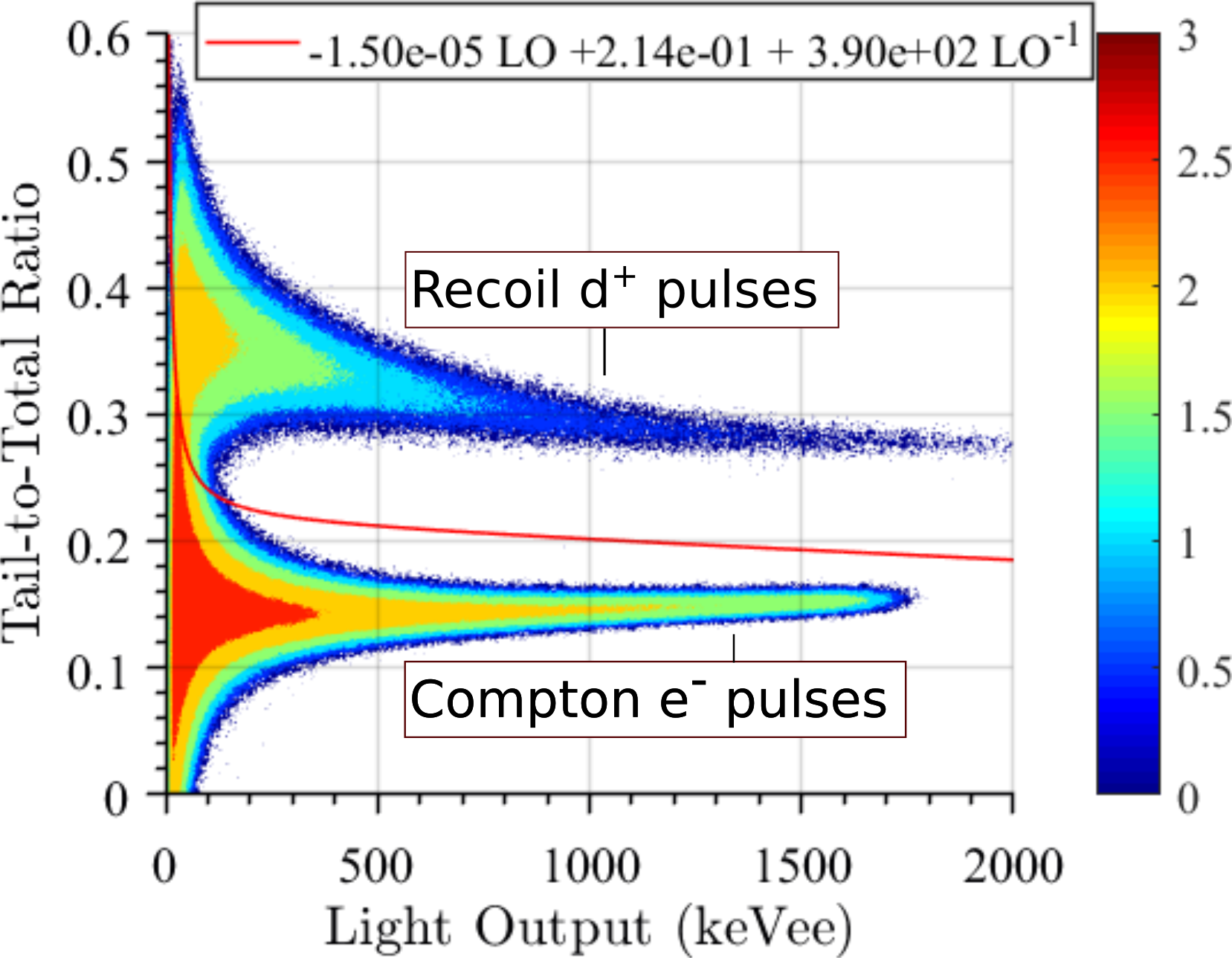}
    \caption{PSD of $^{252}$Cf source measured with the stilbene\-/d$_{12}$ detector. The PSD discrimination equation as a function of pulse light output (LO) was used to select neutron pulses. The colorbar indicates the number of pulses in logarithmic scale.}
    \label{PSD_Cf}
\end{figure}
Fig. \ref{PSD_2plots} shows the PSD scatter-density plots of a $^{239}$PuBe source ($1.3\times10^6$n/s, June 1958) for stilbene\-/d$_{12}$ and $^1$H$\-/$stilbene, using the optimized parameters discussed in Section \ref{PSD}. While both detectors achieved a good PSD, the stilbene\-/d$_{12}$ shows wider separation between the neutron and gamma-ray regions, compared to the $^1$H$\-/$stilbene owing to the higher ionization density of the recoil deuteron. In Fig. \ref{PSD_2plots}, a light-output threshold of 46.7~keVee was used for both detectors. A slice of the $^{239}$PuBe PSD scatter-density plot (Fig. \ref{PSD_2plots}) is shown in Fig. \ref{PuBe_slice} for the 2025-2292 keVee region. The gamma-ray pulses are centered around 0.14 tail-to-total ratio (TTR), and the deuteron recoil pulses are centered around 0.27 TTR. The smaller peak at approximately 0.23 TTR represents proton recoil pulses. While the primary neutron interaction in stilbene-d$_{12}$ is n-d scattering, break-up reactions can also occur and result in protons being detected along with recoil deuterons, as shown in Fig. \ref{PuBe_slice}. The two main reactions that contribute to the production of recoil protons at neutron energies below 20 MeV are d(n,2n~p) and $^{12}$C(n,p)$^{12}$B, with neutron threshold energies of 3.34 MeV and 13.63 MeV, respectively.
The contribution from recoil protons from breakup reactions was not rejected via PSD and was also accounted for in the response matrix simulation to obtain an accurate detector response, described in Section \ref{RM}, and hence a correct unfolded spectrum. 

\subsection{Response Matrix}\label{RM}
\begin{figure}[h!]
    \centering
    \captionsetup{justification=centering}
    \includegraphics[width=100mm]{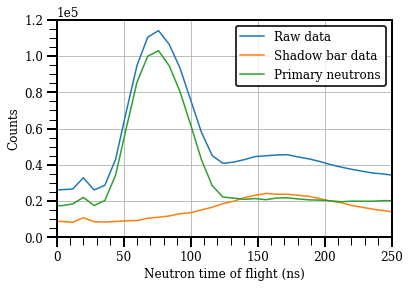}
    \caption{Distribution of neutron flight times for the raw data, the shadow-bar data, and the shadow-bar data subtracted from the raw data, i.e., only the primary neutrons.}
    \label{neutTOFdist}
\end{figure}

We used the PSD line shown in Fig.\ref{PSD_Cf} to discriminate and select neutron pulses detected during the TOF experiment. 
We then calculated the TOF distribution, relative to the trigger detector, for the bare and shadowed cases (Fig. \ref{neutTOFdist}). Fig. \ref{neutTOFdist} shows that the shadow-bar technique is effective in shielding the detector from primary neutrons and selecting only those scattered and detected later in time. The shadow-bar TOF distribution was subtracted from the bare detector measurement and yielded the net TOF, only including the contribution from source neutrons with an uncollided path from the source to the detector.

\begin{figure}[h!]
    \centering
    \captionsetup{justification=centering}
    \includegraphics[width=120mm]{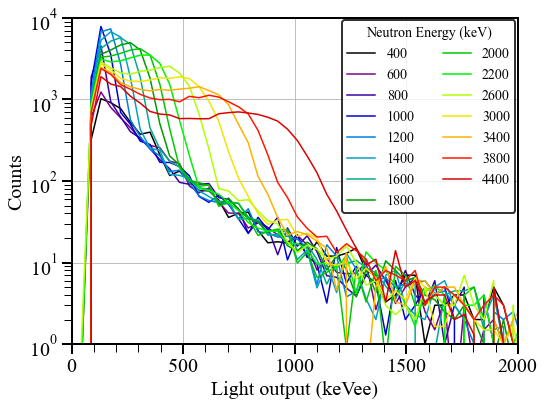}
    \caption{Response matrix of the stilbene\-/d$_{12}$ detector. Each curve is the light-output response to neutrons in a quasi-monenergetic interval.}
    \label{RespMatr}
\end{figure}

We obtained the stilbene\-/d$_{12}$ response matrix to quasi-monoenergetic neutrons 
in 2.2 ns-wide time bins of the net TOF distribution,  shown in Fig. \ref{RespMatr}. Each curve in Fig. \ref{RespMatr} represents the stilbene\-/d$_{12}$ light-output response to the neutrons within each  corresponding energy bin. The light output was calculated as the calibrated pulse integral distribution of the neutron pulses within each time bin. The uncertainty in the measured neutron energy is discussed in the following section and shown in Fig. \ref{LOcurve}. 

\subsection{Light Output Response to Quasi-monoenergetic Neutrons}

The scintillation yield of organic scintillators to nuclear recoils is non-linear and lower than the light yielded by electron recoil interactions, at comparable deposited energies. This light quenching phenomenon, described by several semi-empirical models \cite{BROOKS1979477, AHLEN1977321}, is mostly due to the higher non light-producing heat associated with the atom cascades produced by nuclear recoils, compared to electrons. Therefore, the non-linear light output response to deuteron recoils needs to be characterized using a wide range of deuteron energies. For the specific purpose of this work, we calculated the stilbene\-/d$_{12}$ light output response to quasi-monoenergetic neutrons and used it to simulate the stilbene\-/d$_{12}$ response matrix needed for unfolding in the 0.1 - 20 MeV neutron energy range.
The maximum energy of a nuclear recoil from neutron scattering kinematics is given by Equation \ref{recoilE}, where $A$ is the mass number of the recoil particle. In our case, the maximum possible energy $E_r$ of the recoil deuteron (A=2) is 8/9 the energy of the incident neutron, $E_n$.
 
\begin{equation}\label{recoilE}
    E_r=\frac{4A}{(A+1)^2}E_n
\end{equation}

\begin{figure}[h!]
    \centering
    \captionsetup{justification=centering}
    \includegraphics[width=120mm]{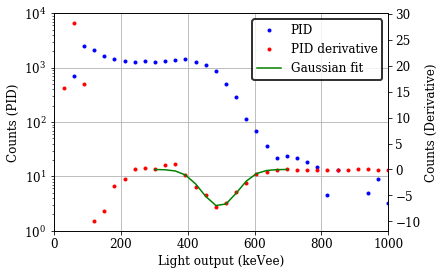}
    \caption{The PID as a function of the light output (LO), its derivative, and the Gaussian fit to find the minimum of the derivative, for the energy bin centered at 3.4 MeV. The equation of this Gaussian fit is $y=-7.11e^{\frac{-(LO-495.84)^2}{2(52.45)^2}}$.}
    \label{Gauss}
\end{figure}

We calculated the minimum of the derivative of each PID measured in TOF mode to determine the light output corresponding to a full energy deposition from single-scattering on deuterium, using the method described by Kornilov et al. \cite{KORNILOV2009226}. A Gaussian distribution was then fitted on the derivatives and we found the light output for a full-energy deposition as the mean of the fitted distribution. Fig. \ref{Gauss} shows an example of a binned PID, its derivative, and the fitted Gaussian distribution in response to 3.4 ± 0.16 MeV neutrons.
The light output response to deuteron recoils was obtained by repeating this procedure for each energy bin. 

\begin{figure}[h!]
    \centering
    \captionsetup{justification=centering}
    \includegraphics[width=120mm]{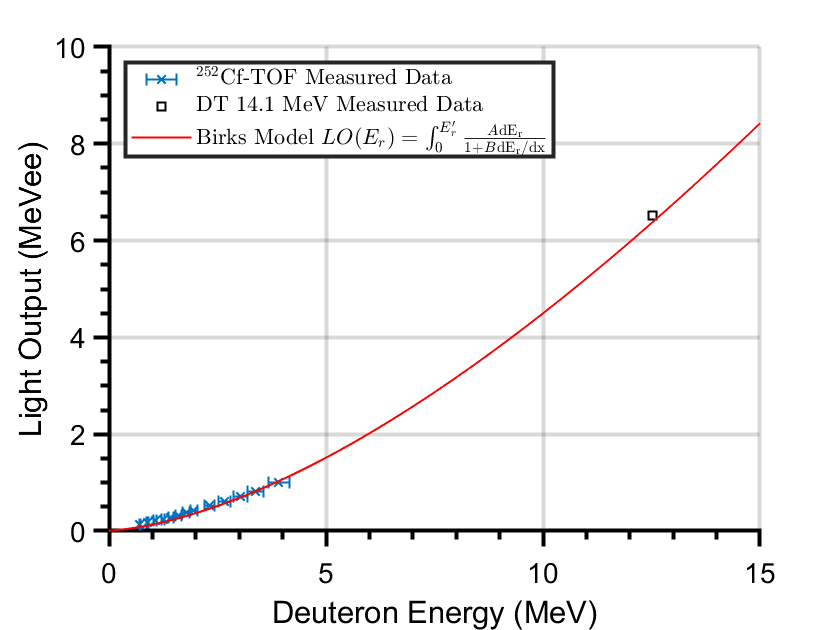}
    \caption{Light output curve of stilbene\-/d$_{12}$ with fitted Birks' model.}
    \label{LOcurve}
\end{figure}

After obtaining the measured light output, we fit the semi-empirical relationship proposed by Birks to the data. The Birks' model \cite{Birks}, given in Equation \ref{BirksEq}, describes the light output functions, including the quenching effect, through the coefficient \textit{B}. In Equation \ref{BirksEq}, $A$ is a conversion coefficient in MeVee/MeV, $E_r$ is the energy deposited by the recoil deuteron in MeV and $dE_r/dx$ is its LET in MeV cm$^2$ mg$^{-1}$.
\begin{equation}\label{BirksEq}
    L(E_r)=\int_{0}^{E'_r} \frac{A\:dE_r}{1+B\:dE_r/dx} 
\end{equation}
Fig. \ref{LOcurve} shows the measured data points and the  fitted model. 
The coefficients that provided the best fit 
are listed in Table \ref{Birks_coeff}. 

\begin{table}[h!]
    \centering
    \begin{tabular}{ |c|c|c| }
    \hline 
    Coefficient & Value & Uncertainty\\
    \hline 
    A (MeVee MeV$^{-1}$) & 2.1 & $\pm$ 0.005\\
    B (mg MeV$^{-1}$ cm$^{-2}$) & 27 & $\pm$ 0.05\\
    \hline 
    \end{tabular}
    \captionsetup{justification=centering}
    \caption{Coefficients of the Birks model fit and their relative uncertainties. }
    \label{Birks_coeff}
\end{table}


\subsection{Simulated Response Matrix for Regular and Deuterated Stilbene}\label{respMatr}

\begin{figure}[h!]
    \centering
    \captionsetup{justification=centering}
    \includegraphics[width=120mm]{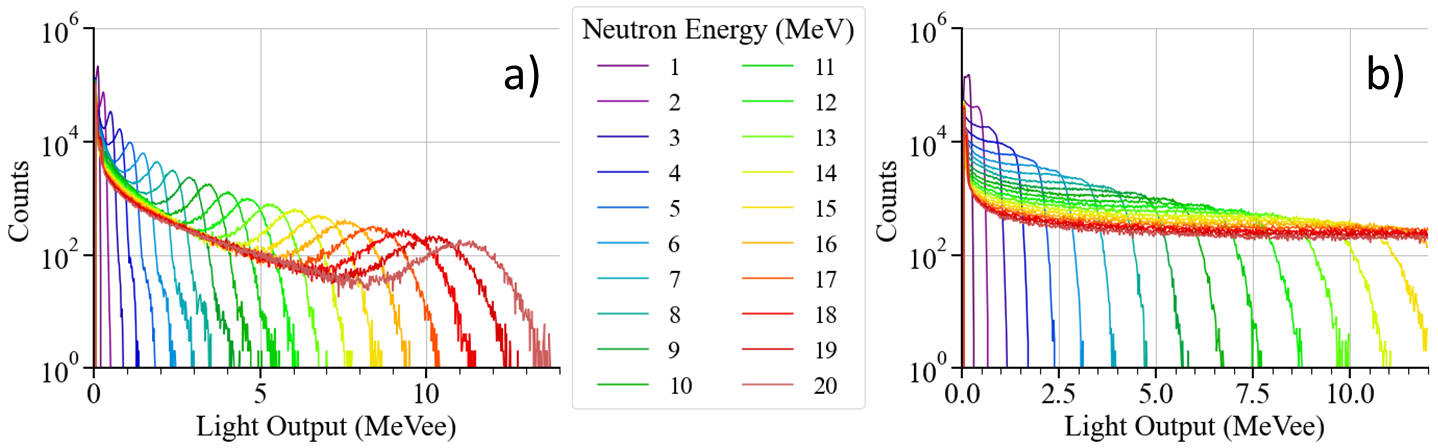}
    \caption{Simulated response matrix of (a) stilbene\-/d$_{12}$ and (b) $^1$H$\-/$stilbene on a logarithmic scale, using MCNPX-PoliMi.}
    \label{regSBSimRM}
\end{figure}

We used the light-output response function shown in Fig. \ref{LOcurve} to simulate the response of stilbene\-/d$_{12}$ to monoenergetic neutron beams with energies ranging from 0.1 to 20 MeV in MCNPX-PoliMi with MPPost \cite{PoliMi}. Breakup reactions in the scintillator were also included in the simulation and the light output from proton recoils produced by these reactions was simulated using the specific model for proton recoils in $^{1}$H-stilbene, described below. The response matrix of $^{1}$H-stilbene was also simulated, using the light output model by Prasad et al. \cite{Prasad}. Fig. \ref{regSBSimRM} shows the simulated response matrices. Stilbene\-/d$_{12}$'s response exhibits a peak-like structure at the maximum energy deposited, which is absent in the $^1$H$\-/$stilbene's response. 

\subsection{Unfolding Results}
\begin{figure}[h!]
    \centering
    \captionsetup{justification=centering}
    \includegraphics[width=110mm]{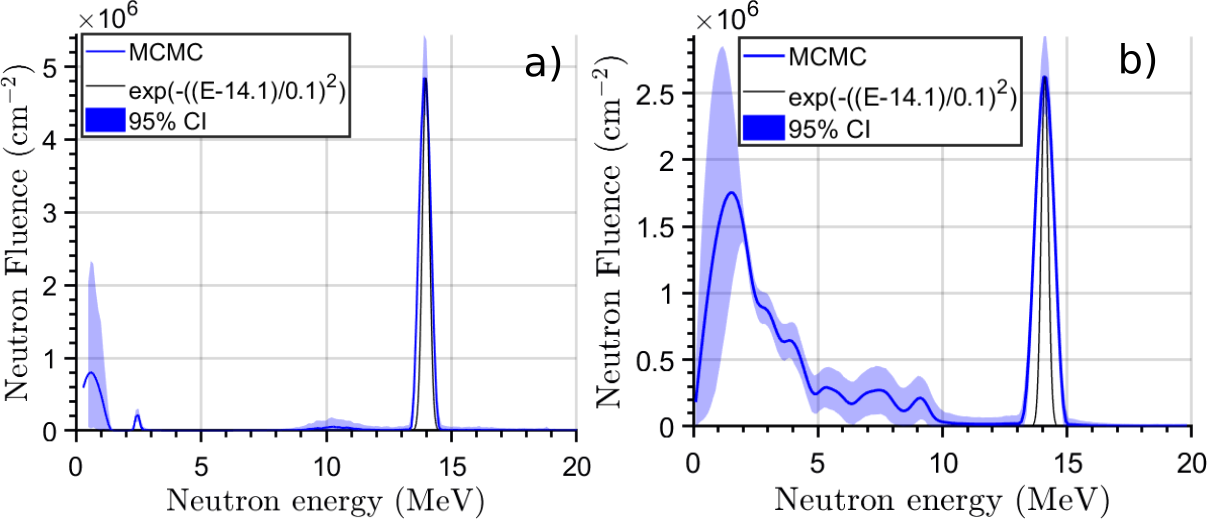}
    \caption{Results from unfolding the spectrum of a DT fusion source for (a) stilbene\-/d$_{12}$ and (b) $^1$H$\-/$stilbene. The reference spectrum is normalized to the maximum of the distributions and shown for comparison purposes.}
    \label{UF_DT}
\end{figure}

\begin{figure}[h!]
    \centering
    \captionsetup{justification=centering}
    \includegraphics[width=110mm]{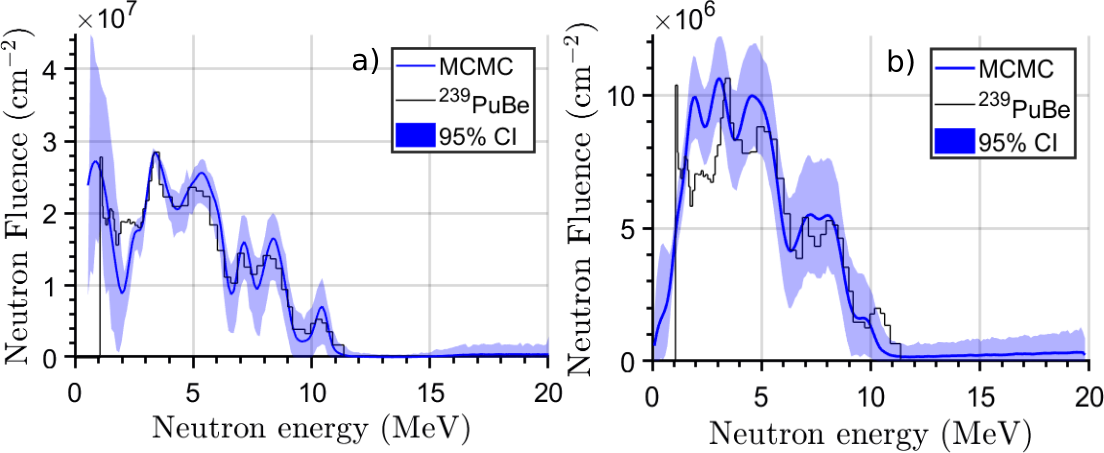}
    \caption{Results from unfolding the spectrum of a $^{239}$PuBe fusion source for stilbene\-/d$_{12}$ (a) and $^1$H$\-/$stilbene (b). The reference spectrum is normalized to the maximum of the distributions and shown for comparison purposes.}
    \label{UF_PuBe}
\end{figure}

We have measured the response of stilbene\-/d$_{12}$ and $^1$H$\-/$stilbene to a 14.1 MeV D-T neutron beam (Thermo Fisher Scientific P211, Fig. \ref{UF_DT} a) and a $^{239}$PuBe neutron source (Fig. \ref{UF_PuBe}), performed PSD, selected neutron pulses and derived the light output spectra for both sources and detectors. We then derived the source neutron spectra using the methods described in Section \ref{unfolding}. The unfolding algorithm is based on a response matrix in units of counts per unit fluence, and therefore, it allows a direct measurement of the impinging fluence. Qualitatively, as shown in Fig. \ref{UF_DT} and \ref{UF_PuBe}, the neutron spectra measured using the stilbene\-/d$_{12}$ detector resemble better the reference ground truth spectra and shows narrower credible intervals (CI) when compared to the unfolded $^1$H$\-/$stilbene response. 

For the $^{239}$PuBe, the SAM values between the unfolded spectra and the reference one are 11$\degree$ and 9$\degree$, for $^1$H$\-/$stilbene and stilbene\-/d$_{12}$, respectively, while for the DT neutrons, the SAM measures 37$\degree$ and 19$\degree$ for $^1$H$\-/$stilbene and stilbene\-/d$_{12}$, respectively. A higher SAM value in the latter case for both detectors is due to the larger width of the peak at 14.1 MeV, compared to the reference spectrum. As reference DT spectrum, we considered a Gaussian distribution with an average energy of 14.1 MeV and a width of 100 keV; the reference $^{239}$PuBe spectrum is by Anderson and Neff \cite{ANDERSON1972231}.
In Fig.\ref{UF_DT}a, one can observe the main neutron energy at 14.1 MeV and a small peak at 2.5 MeV, with a small associated uncertainty. According to the manufacturer \cite{Simpson}, the generator (P211 by Thermo Fisher Scientific) beam is a mixture of deuteron and triton ions, in monatomic and diatomic forms. Therefore, a variety of ions can hit the target, which contains deuterium and tritium. Neutrons can thus be produced through d+t, d+d, and t+t reactions. The output is expected to be mainly due to the d+t reaction (97-98\%) and the remaining to d+d and t+t reactions, because of the typical cross section ratios and target composition. These fractions are consistent with the relative areas under the measured peaks.

\section{Conclusions}
We have characterized the neutron light output response of a 32~cm$^3$ stilbene\-/d$_{12}$ detector and confirmed the excellent pulse shape discrimination capability in the 0.5–10 MeV neutron energy range. The stilbene\-/d$_{12}$ response to quasi-monoenergetic neutrons exhibits distinct, broad peaks corresponding to the incident neutron energy, with the energy deposited by the deuteron recoil $E_d = 8/9 \times E_n$. The stilbene\-/d$_{12}$ light output spectrum hence can provide neutron energy information from the recoil deuteron spectrum without TOF. Based on the experimental light output, the response matrices of stilbene\-/d$_{12}$ and $^1$H$\-/$stilbene were simulated using MCNPX-PoliMi for a wide neutron energy range (0.1 - 20 MeV). We then used the simulated response matrices to compare the spectroscopy capability of the two detectors to reconstruct both monoenergetic (14.1 MeV DT) and continuous ($^{239}$PuBe) spectra.
We formulated the problem of deconvolving the energy of interacting neutrons from the light output response in a Bayesian framework, where we added prior information in terms of a smoothing parameter. A Bayesian unfolding algorithm was used to derive the spectral fluence and estimate its uncertainty through MCMC sampling.
The spectroscopy unfolding did not require any additional prior information on the interacting neutron spectrum. When compared to a $^1$H$\-/$stilbene of similar size and analogous production process, the spectra derived from the unfolded stilbene\-/d$_{12}$ light output response more accurately resembled the reference spectra. These results confirm that stilbene\-/d$_{12}$ scintillators are suitable for neutron spectroscopy without the need of time of flight. The excellent PSD and neutron spectroscopy capabilities of stilbene\-/d$_{12}$ make it suitable for many applications including nuclear reaction studies, radiation protection, nuclear security, and non-proliferation, among others. The results reported in this work are currently being extended in additional experiments with a larger 140~cm$^3$ stilbene\-/d$_{12}$ crystal. 


\section*{Acknowledgements}
This work was funded in part by the Nuclear Regulatory Commission (NRC) Faculty Development Grant 31310019M0011, and NRC fellowship grants NRC-HQ841560020 and NRC-31310018M0029. Crystal growth at LLNL was supported by the DOE/NA-22 office. This material is based upon work supported by the U.S. Department of Energy, Office of Science, Office of Nuclear Physics, under Award Number DE-AC05-00OR22725

\newpage
\bibliography{main.bib}

\end{document}